# *In vitro* AND *in vivo* ANTILEISHMANIAL EFFICACY OF A COMBINATION THERAPY OF DIMINAZENE AND ARTESUNATE AGAINST *Leishmania donovani* IN BALB/C MICE

Joshua Muli MUTISO(1,3), John Chege MACHARIA(1), Mustafa BARASA(3), Evans TARACHA(1), Alain J. BOURDICHON(2) & Michael M. GICHERU(3)

## SUMMARY

The *in vitro* and *in vivo* activity of diminazene (Dim), artesunate (Art) and combination of Dim and Art (Dim-Art) against *Leishmania donovani* was compared to reference drug; amphotericin B. $IC_{50}$ of Dim-Art was found to be 2.28 ± 0.24 µg/mL while those of Dim and Art were 9.16 ± 0.3 µg/mL and 4.64 ± 0.48 µg/mL respectively. The $IC_{50}$ for Amphot B was 0.16 ± 0.32 µg/mL against stationary-phase promastigotes. *In vivo* evaluation in the *L. donovani* BALB/c mice model indicated that treatments with the combined drug therapy at doses of 12.5 mg/kg for 28 consecutive days significantly ($p < 0.001$) reduced parasite burden in the spleen as compared to the single drug treatments given at the same dosages. Although parasite burden was slightly lower ($p < 0.05$) in the Amphot B group than in the Dim-Art treatment group, the present study demonstrates the positive advantage and the potential use of the combined therapy of Dim-Art over the constituent drugs, Dim or Art when used alone. Further evaluation is recommended to determine the most efficacious combination ratio of the two compounds.

## INTRODUCTION

Visceral leishmaniasis (VL), also known as Kala azar, is a protozoan systemic infection which is always fatal if left untreated. This illness was included by the World Health Organization in the list of neglected tropical diseases targeted for elimination by 2015 (MALTEZOU, 2010). The first line chemotherapeutic agents, such as pentavalent antimonial compounds: pentostam and glucantime (CROFT & COOMBS, 2003), remain the drugs of choice for all forms of leishmaniasis in spite of their reported toxicity. Amphotericin B (Amphot B) and pentamidine, the second line drugs, remain of limited value because of toxicity (BERMAN *et al.*, 1998) and prohibitively high cost (WHO, 2007). Miltefosine, initially developed as an anticancer drug, is the first effective oral treatment of visceral leishmaniasis (VL) and the latest antileishmanial drug to enter the market (Croft and Coombs, 2003). Drug resistance in leishmaniasis has been reported in all these drugs (MAAROUF *et al.*, 1998; CROFT *et al.*, 2006). Even before miltefosine is introduced into the market or into control programs, preliminary data from a phase IV trial in India involving domiciliary treatment with miltefosine and weekly supervision suggests doubling of the relapse rate (SUNDAR & MURRAY, 2005); this provides warning that drug resistance could develop quickly and plans are required to prevent it.

Combination therapy of antileishmanial drugs is currently considered as one of the most rational approaches to lower treatment failure rate and limit the spreading of drug resistance (GUERIN *et al.*, 2002; GAZANION *et al.*, 2011). Diminazene (Dim) has been tested and found to have trypanocidal and leishmanicidal activities on various strains of trypanosomes (*Trypanosoma vivax, T. evansi* and *T. congolense*) and on *L. donovani* in hamsters (CLEMENT *et al.*, 1992).

Studies conducted in the 1970s suggested the potential of berenil (diminazene diaceturate) as leishmanicidal agent (LEON *et al.*, 1977). Trypan®: a diminazene drug was tested against both *Leishmania major* and *L. donovani in vitro* by MACHARIA *et al.* (2004) and the drug was found to be efficacious. The same experiment also carried out *in vivo* studies with Trypan® using BALB/c mice with cutaneous leishmaniasis after infection with *L. major* promastigotes. Results showed delayed lesion development depending on the route of drug administration. Dim granules represent a new formulation, containing, in part, Dim aceturate in addition to other new ingredients. Dim is formulated for both oral and parenteral (intramuscular and intravenous) administration. Artesunate (Art), dihydroartemisinin-10α- hemisuccinate, is a semi-synthetic derivative of artemisinin which is used for the treatment of both uncomplicated and severe malaria (OLLIARO *et al.*, 2001). It is formulated for oral, parenteral and rectal administration and used clinically worldwide. Because of rapid hydrolysis to dihydroartemisinin (also referred to as artenimol), Art is considered by many as a prodrug of the latter (NAVARATNAM *et al.*, 2000). Dim and Art granules are relatively new products and their synergistic effect has not been tested for its potential application against leishmaniasis. The present study investigated the protective potential of Dim combined with Art in BALB/c mice experimentally infected with *L. donovani*. The efficacy of the drug combination and the survivorship of the study subjects are discussed.

---
(1) Department of Tropical and Infectious Diseases, Institute of Primate Research, P.O. Box 24481 - 00502 Karen, Nairobi, Kenya.
(2) BPM Bulk Medicine and Pharmaceuticals Production, TROPMED GMbH, Neuhofer Welche 48, D - 19370 Parchtm, Germany.
(3) Department of Zoological Sciences, School of Pure and Applied Sciences, Kenyatta University, P.O. Box 43844 - 00100 Nairobi, Kenya.
**Correspondence to:** Joshua Muli Mutiso, Tel.: + 254 202 161 157, Fax: +254 202 606 231. E-mail: mutisovic@yahoo.co.uk

## MATERIALS AND METHODS

**Parasites for infection:** *Leishmania donovani* strain NLB-065 originated from the spleen of an infected patient and is maintained by intracardiac hamster-to-hamster passage at the Institute of Primate Research. A hamster splenic aspirate was cultured in Schneider´s Drosophila insect medium supplemented with 20% fetal bovine serum and 100 µg/mL of gentamicin at 25 °C till stationary phase. Parasites harvested in stationary phase after 8-10 days of culture were centrifuged at 2500 rpm (Servoll 6000D) for 15 min at 4 °C and washed three times in sterile phosphate-buffered saline before being counted and used for inoculation in animals.

**Chemicals:** Dim diaceturate and Art (dihydroartemisinin-10α-hemisuccinate) granules were provided by Dr Alain Bourdichon (TropMed, Germany). The compounds were weighed separately, mixed in a 1:1 ration by weight and used in dosages of 12.5 mg/kg of body weight. Art and Dim were also individually used at dosages of 12.5 mg/kg. Amphot B at doses of 12.5 mg/kg was used as a positive control drug.

***In vitro* evaluation against *Leishmania donovani* promastigotes:** Stationary-phase promastigotes harvested as described above were counted and suspended in a concentration of $2.0 \times 10^6$ parasites/mL in culture medium. With a few modifications, the tests were performed as previously described (OKPEKON *et al.*, 2004). The tests were performed in 96-well microtitre plates maintained at 26 ºC under 5% $CO_2$ atmosphere. Two hundred microliters of complete Schneider´s Drosophila medium was placed in the wells containing the maximum concentrations of the compounds and 100 µL in the next wells (2 to 12) and controls; 2 µL of compound solutions of 20 mg/mL in distilled water were added to wells number 1 and serial dilutions (ranging from 100 µg/mL to 0.049 µg/mL) in the wells were performed. Hundred microliters of culture medium containing $2.0 \times 10^6$ stationary-phase *Leishmania* parasites/mL was added to each test well. Tests were performed two times each with three replications for each test compound concentration. Parasite observations and counting were done using a microscope. The results are expressed as the concentration inhibiting parasite growth by 50% ($IC_{50} \pm SD$) after 72 h incubation period. The initial concentrations for testing were 100 µg/mL.

***In vivo* evaluation against *Leishmania donovani*:** Six to eight week-old BALB/c mice of mixed sexes were infected with $1 \times 10^6$ virulent *L. donovani* strain NLB-065 harvested at stationary phase as previously described. Infected mice were kept for nine weeks for symptomatic establishment of VL. Infection was assessed in three mice by culture of splenic aspirate and impression smears. The animals were then divided into five groups of six mice each and treated with either Amphot B, Dim, Art, or Dim-Art. One group was not treated and it served as a control. All drugs were given at dosages of 12.5 mg/kg of body weight. All doses were administered consecutively for 28 days from week 10 post infection. During the treatment period, parasite load was determined in any animal that died. In week 14 post infection (day 97), all mice groups were sacrificed and parasite numbers determined microscopically by counting the number of amastigotes/500 splenic cell nuclei in Giemsa-stained impression smears. Amastigote burden was compared for both treated and untreated mice groups.

**Mice survival rate:** During treatment period, weeks 10 to 14, all animal groups were observed at least every 12 hours and any death that occurred during this time was recorded.

**Statistical analysis:** All parasite burden data were expressed as the mean per 500 cell nuclei of spleen cells ± standard deviation. Differences among groups were analyzed by one-way analysis of variance (ANOVA), and the post hoc Tukey-Krammer test method was used for multiple comparisons. A *p*-value of $p < 0.05$ was considered statistically significant. All analyses were performed using the GraphPad InStat software.

## RESULTS

The *in vitro* antileishmanial effects of Dim, Art, Dim-Art and reference drug Amphot B on the growth of *L. donovani* promastigotes are presented in Table 1. These results indicate that Dim-Art ($IC_{50}$ of 2.28) was more than twice and more than four times effective in inhibiting promastigotes growth as compared to Art ($IC_{50}$ of 4.64) and Dim ($IC_{50}$ of 9.16) respectively.

**Table 1**
$IC_{50}$ values of Dim, Art, Dim-Art and reference drug Amphot B on *L. donovani* promastigotes

| Compound | $IC_{50}$ (µg/mL) |
| --- | --- |
| Dim | 9.16 ± 0.3 |
| Art | 4.64 ± 0.48 |
| Dim-Art | 2.28 ± 0.24 |
| Amphot B | 0.16 ± 0.32 |

Mean of $IC_{50}$ (µg/mL) ± SD

The *in vivo* antileishmanial effects of 12.5 mg/kg of each of Dim, Art and Dim-Art and reference drug, Amphot B are presented in Figure 1. These results indicate that, Dim/Art significantly reduced parasite loads when used in this combined form than when used as single drug therapy of Dim-Art ($p < 0.001$).

The reference drug, Amphot B, indicated only a slight reduction of parasite load when compared to the Dim-Art ($p < 0.05$). There was no difference in antleishmanial activity by Dim and Art ($p > 0.05$). However, all drugs significantly reduced parasite loads as compared to the control group ($p < 0.001$).

The survival rates of mice during treatment are shown in Figure 2. No death was observed in any mouse from each of the Art, Dim-Art or the Amphot B groups throughout the experimental period and this concluded a highly significant survival rate in these groups as compared to the control group ($p < 0.001$). The rate of mice survival in the Dim group was not different from all the other treated mice groups ($p > 0.05$). This survival rate was however significantly higher in the Dim treated group than in the control mice group ($p < 0.05$).

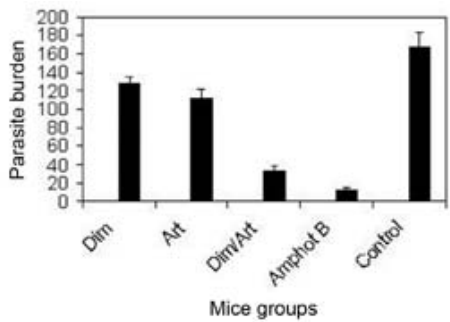

**Fig. 1** - *In vivo* efficacy of diminazene (Dim) alone, Artesunate (Art) alone, combination therapy of diminazene/artesunate (Dim-Art) and amphotericin B (Amphot B) in BALB/c mice infectecd with *Leishmania donovani*. The drugs were given at dosages of 12.5 mg/kg for 28 consecutive days following nine weeks of disease. At the time of death of mice during treatment or at the time of termination at week 14, splenic parasite loads were determined and compared for all the groups. Data shown represents the mean ± SD for each group.

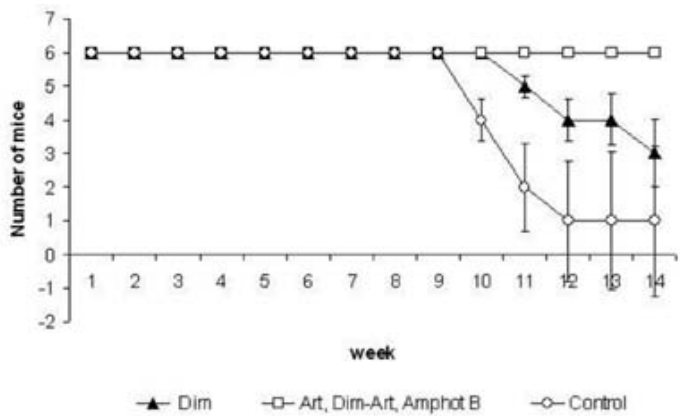

**Fig. 2** - Survivorship of *Leishmania donovani* infected-BALB/c mice during treatment with diminazene (Dim), artesunate (Art), diminazene plus artesunate (Dim-Art) or amphotericin B (Amphot B). BALB/c mice (n = 6) were infected with *L. donovani* and 10 weeks later, treated daily for 28 consecutive days. Data show surviving number of mice per group ± SD recorded every week for 14 weeks.

## DISCUSSION

Drugs currently available for leishmaniasis treatment often show parasite resistance, high toxic side effects and prohibitive costs commonly incompatible with patients from the tropical endemic countries (SIQUEIRA-NETO *et al.*, 2010). Combination therapy between commercially available drugs that are aimed to reduce cost, toxicity and duration of treatments, represents a promising rational alternative (GAZANION *et al.*, 2011). In this study, we evaluated the possible application of Art, an antimalarial drug, in combination with Dim diaceturate as an antileishmanial drug therapy. Dim has been tested and found to have trypanocidal and leishmanicidal activities on various laboratory strains of trypanosomes (*Trypanosoma vivax, T. evansi* and *T. congolense*) and on *L. major* in mice (MACHARIA *et al.*, 2004) and *L. donovani* in hamsters (CLEMENT *et al.*, 1992). Dim diaceturate is a Dim currently undergoing trial for treatment of trypanosomiasis. Our present study shows the positive advantage of Dim-Art combination over the use of its constituent drug compounds in the treatment of visceral leishmaniasis. Dim-Art indicated an efficacy level of more than twice and more than four times in inhibiting parasites growth in culture as compared to Art and Dim respectively. This is an indication that the combined therapy is more potent than the single drugs with the presence of Art contributing more to the potency than the Dim compound. The lack of complete elimination of parasites by the Dim-Art combined therapy or the Amphot B may be attributed to delayed start of treatments of infected mice. Furthermore, treatment began when most of the mice had shown severe symptoms of the disease and it may be possible that these two drugs could only completely eliminate parasites when administered for a longer period in cases of advanced disease.

In the current study, we monitored for animal survival during the experimental period. The survival of 100% of the populations treated with either, artesunate, Dim-Art or Amphot B up to the end of the experimental period may be attributed to significantly reduced parasite loads observed in these groups. This is an indication of a potential value of these drug compounds in the control of *Leishmania* parasite multiplication. Fifty percent of the total population treated with Dim and over 80% of mice in the control group died before termination of the experiment. It was observed that any animal that died from any of these two groups before the end of the experimental period had a parasite load of at least 130 parasites per 500 nucleated spleen cells. BALB/c mice fail to control infection and they develop progressive lesions and systemic disease (SACKS & NOBEN-TRAUTH, 2002). This may point to inability of the Dim drug to control parasite growth and hence the death of some mice in this group. Following administration of any of the compounds used in this study, we did not observe any side effects attributable to the dosages of the regimen given. Furthermore, earlier experiments in our laboratory showed no side effects when Dim was administered in *L. major*-infected mice (MACHARIA *et al.*, 2004). We did not observe any adverse effects of these chemicals when delivered into the mice nor did we observe any morphological changes in the parasites in the *in vitro* test system and this may give an indication that parasite death or the death of mice was not caused by drug toxicity.

In conclusion, we have demonstrated the positive advantage and possible application of Dim-Art drug combination in the safe and effective treatment of VL. It is possible that much of the efficacy of this drug combination is attributable to the presence of the Art compound and hence the need for further experimental design to establish the curative combination ratio and toxicity parameters of these compounds.

## RESUMO

### Estudo *in vitro* e *in vivo* da eficácia anti leishmaniótica de terapêutica combinada de Diminazene e Artesunate contra *Leishmania donovani* em camundongos Balb/c

A atividade *in vitro* e *in vivo* de Diminazene (Dim), Artezunate (Art) e a combinação Dim e Art (Dim-Art) contra *Leishmania donovani* foi comparada com a droga de referência Anfotericina B. IC$_{50}$ da Dim-Art foi 2,28 ± 0,24 µg/mL enquanto aquelas de Dim e Art foram 9,16 ± 0,3 µg/mL e 4,64 ± 0,48 µg/mL respectivamente. O IC$_{50}$ da Anfotericina B foi 0,16 ± 0,32 µg/mL contra a fase estacionária de promastigotas. A avaliação *in vivo* do modelo de *L. donovani* em camundongos Balb/c indicou que os tratamentos com a terapêutica de drogas combinadas em doses de 12,5 mg/kg por 28 dias consecutivos significantemente (p < 0,001) reduziu a carga parasitária no baço quando comparada a tratamentos com uma única droga dada nas mesmas dosagens. Embora a carga parasitária tenha sido levemente mais baixa (p < 0.05) no grupo Anfotericina B quando comparada com o grupo tratado Dim-Art, o estudo presente demonstra a vantagem positiva do uso potencial da terapêutica combinada Dim-Art sobre drogas como Dim ou Art quando usadas isoladamente. Posterior avaliação é recomendada para determinar a média de combinação mais eficaz dos dois compostos.